\documentclass[12pt]{iopart}
\newtheorem{theorem}{THEOREM}
\renewcommand{\theequation}{\Roman{section}.\arabic{equation}}

\begin{document}

\title{On the Derivation of Conserved Quantities in Classical Mechanics}
\author{Paulus C.\ Tjiang and
Sylvia H. Sutanto}
\address{Department of Physics, Faculty of Mathematics and Natural Sciences \\
Universitas Katolik Parahyangan, Bandung 40141 - INDONESIA}

\ead{\mailto{pctjiang@home.unpar.ac.id},
\mailto{sylvia@home.unpar.ac.id}}

\begin{abstract}
Using a theorem of partial differential equations, we present a
general way of deriving the conserved quantities associated with a
given classical point mechanical system, denoted by its
Hamiltonian. Some simple examples are given to demonstrate the
validity of the formulation.
\end{abstract}

\pacs{02.60.Lj, 45.20.Jj}

\section{Introduction.}
\label{introduction}

It has been well-known for many years of the history of physics
that symmetry principles and conservation laws play a crucial role
in understanding physical
phenomena~\cite{Sudarshan74,Itzykson80,Weinberg95}. Many physical
problems can be solved significantly easier by taking into account
symmetry consideration, and most importantly, one can gain deep
insight of the nature of the phenomena. Many fields of study in
modern physics developed rapidly through the understanding of
symmetry principles, for instance, the gauge symmetries in quantum
field theory and elementary particle
physics~\cite{Abers73,Cheng84,Moriyasu85}.

The existence of symmetries in a physical system means that there
exists one or more transformations that leave the physical system
unaltered. Since physical systems can be completely described by
their Lagrangians / Hamiltonians, it is therefore natural to
expect the symmetry transformations to be canonical ones. The
number of symmetry transformations that can be generated from the
system depends on the number of conserved quantities of the
associated system. It has been well-known that the conserved
quantities of a physical system relate closely to the generator of
the associated symmetry
transformations~\cite{Sudarshan74,Itzykson80,Goldstein80}.

In this paper, we shall present a derivation of all possible
conserved quantities associated with a given physical system
directly from its associated conservation laws. We shall limit our
task on classical point mechanics, although it may be possible to
extend the derivation techniques to quantum mechanics and field
theories, once the connection between classical and quantum
mechanics is established via canonical
realizations~\cite{Sudarshan74,Pauri66}. Some simple examples
associated with time-independent and time-dependent systems will
be considered.

The paper is organized as follows : a brief overview of classical
mechanics is given in Section~\ref{classical}, in which we shall
use Lagrangian and Hamiltonian approaches. This includes a brief
review of canonical transformations, especially a subset of the
transformations called infinitesimal canonical transformations.
The relation between symmetry canonical transformations and
conservation laws, as well as the use of a theorem of partial
differential equations to obtain the conserved quantities are
given in Section~\ref{symmetry}. In
Section~\ref{time-independent}, we shall discuss the derivation of
conserved quantities associated with time-independent systems. A
similar discussion for time-dependent cases is given in
Section~\ref{time-dependent}. Some simple examples are used to
demonstrate the validity of the formulation. Section~\ref{summary}
presents summary and conclusion of the discussion, as well as a
brief remark of the extension possibility of the techniques to
quantum mechanics and field theories.

\section{Classical Mechanics : Canonical Formalism.}
\label{classical}

A classical system can always be represented by an associated
Lagrangian $L(q_i,\dot{q}_i,t)$ that satisfies
\begin{equation}
\delta A \equiv \delta \int_{t_1}^{t_2} L(q_i,\dot{q}_i,t) dt = 0
\label{action}
\end{equation}
and leads to Euler-Lagrange equations of motions
\begin{equation}
\frac{\partial L}{\partial q_i} - \frac{d}{dt}
\left(\frac{\partial L}{\partial \dot{q}_i} \right) = 0,
\label{E-L}
\end{equation}
where $i = 1 ... N$, {\it N} = number of degrees of freedom.

Defining the conjugate momenta $p_i$ associated with the canonical
coordinates $q_i$ as
\begin{equation}
p_i \equiv \frac{\partial L}{\partial \dot{q}_i}, \label{momenta}
\end{equation}
we may construct a new function $H(q_i,p_i,t)$, the Hamiltonian,
that can equivalently represent the physical system as
$L(q_i,\dot{q}_i,t)$ via a Legendre
transformation~\cite{Goldstein80} :
\begin{equation}
H(q_i,p_i,t) \equiv \sum_j p_j \dot{q}_j - L(q_i,\dot{q}_i,t).
\label{Hamiltonian}
\end{equation}
$H(q_i,p_i,t)$ is known as Hamiltonian. Using Eqs.~(\ref{E-L}) and
(\ref{Hamiltonian}), a set of equations of motion equivalent to
Eq.~(\ref{E-L}), the so-called Hamilton's equations of motion, can
be obtained as follows :
\begin{eqnarray}
\dot{q}_i & = & \frac{\partial H}{\partial p_i}, \nonumber \\
\dot{p}_i & = & -\frac{\partial H}{\partial q_i}, \label{Hamilton-equation} \\
\frac{\partial H}{\partial t} & = & -\frac{\partial L}{\partial
t}. \nonumber
\end{eqnarray}

A set of invertible phase space transformations
\begin{equation}
(q_i,p_i,t) \Longleftrightarrow (Q_i,P_i,t) \label{transformation}
\end{equation}
which preserves Eq.~(\ref{Hamilton-equation}) is called canonical
transformations. It is easy to show that the set of canonical
transformations forms a non-abelian group~\cite{Sudarshan74}, and
that the transformations can be generated by one of the four types
of generating functions $F_1 (q_i,Q_i,t), F_2 (q_i,P_i,t), F_3
(p_i,Q_i,t), F_4 (p_i,P_i,t)$~\cite{Goldstein80}. The old
Hamiltonian $H(q_i,p_i,t)$ and the new counterpart $K(Q_i,P_i,t)$
are related by
\begin{equation}
K(Q_i,P_i,t) = H(q_i,p_i,t) + \frac{\partial F_\alpha}{\partial
t}, \, \, \, \, \, \alpha = 1, ..., 4.
\label{Hamiltonian-relation}
\end{equation}
In the case of a time-independent system, the {\it value} of the
old and new Hamiltonian are the same, even if they have different
{\it functional forms}. In general, $K(Q_i,P_i,t)$ and
$H(q_i,p_i,t)$ have different functional forms, leading to
different physical interpretations.

It is more convenient to use the Poisson bracket formulation,
defined as follows for $A(q_i,p_i,t)$ and $B(q_i,p_i,t)$ :
\begin{equation}
\left\{A,B \right\} \equiv \sum_{i=1}^{N} \left(\frac{\partial
A}{\partial q_i} \frac{\partial B}{\partial p_i} - \frac{\partial
A}{\partial p_i} \frac{\partial B}{\partial q_i} \right).
\label{Poisson-bracket}
\end{equation}
The first two equations in Eq.~(\ref{Hamilton-equation}) can be
written as
\begin{eqnarray}
\dot{q}_i & = & \{q_i,H\} \nonumber \\
\dot{p}_i & = & \{p_i,H\}. \label{Hamilton-equation-2}
\end{eqnarray}
In the case of canonical transformations, it is easy to show that
the necessary and sufficient condition for the transformation
(\ref{transformation}) to be canonical is
\begin{equation}
\{Q_i,P_j\} = \{q_i,p_j\} = \delta_{ij}.
\label{canonical-condition}
\end{equation}

We are now focusing our attention to a subset of canonical
transformations that lies in the neighborhood of the identity
transformation, namely the {\it infinitesimal} canonical
transformations. The transformation can be generated by the
following generating function~\cite{Goldstein80}
\begin{equation}
F_2 (q_i,P_i,t) = \sum_j \left(q_j P_j + \epsilon_j f_j(q_i,p_i,t)
\right), \label{generating-infinitesimal}
\end{equation}
where $f_j (q_i,p_i,t)$ are the generators of infinitesimal
canonical transformations, and $\epsilon_j$ are infinitesimal
parameters. The transformation generated by
(\ref{generating-infinitesimal}) can be written as
\begin{eqnarray}
Q_i & = & q_i + \sum_j \epsilon_j \{q_i,f_j\}, \nonumber \\
P_i & = & p_i + \sum_j \epsilon_j \{p_i,f_j\}.
\label{infinitesimal-transformation}
\end{eqnarray}
It can be shown that the set of infinitesimal canonical
transformations forms an abelian group, and hence it can be used
to form a linear vector space and an algebra~\cite{Sutanto91}. The
algebra of infinitesimal canonical transformation satisfies the
following Poisson bracket relations~\cite{Sudarshan74,Pauri66} :
\begin{equation}
\{f_i, f_j\} = \sum_k C_{ij}^k f_k + d_{ij}
\label{infinitesimal-algebra}
\end{equation}
where $C_{ij}^k$ and $d_{ij}$ are constants depending on the
nature of the transformation. In some transformations, the
constant $d_{ij}$ may be removed by a suitable
substitution~\cite{Sudarshan74,Pauri66}. The relations
(\ref{infinitesimal-algebra}) are closely related to the
commutation relations in Lie algebras~\cite{Gilmore74}.

\section{Symmetries, Conservation Laws and General Conserved Quantities.}
\label{symmetry}

The infinitesimal canonical
transformation~(\ref{infinitesimal-transformation}) that leaves
the Hamiltonian of a given physical system invariant is called the
symmetry canonical transformation of the associated system. The
term {\it invariant} means that the nature of the physical system
is unaltered by the transformation, which means that the old and
transformed Hamiltonians have to be of the same {\it functional
forms}. As a consequence, Eq.~(\ref{Hamiltonian-relation}) can be
written in the form of
\begin{equation}
H(Q_i,P_i,t) = H(q_i,p_i,t) + \frac{\partial F_2}{\partial t},
\label{Hamiltonian-relation-2}
\end{equation}
Inserting Eqs.~(\ref{generating-infinitesimal}) and
(\ref{infinitesimal-transformation}) into
Eq.~(\ref{Hamiltonian-relation-2}) and applying the Taylor
expansion on $H(Q_i,P_i,t)$ to the first order of $\epsilon_i$
together with Eqs.~(\ref{Poisson-bracket}) and
(\ref{Hamilton-equation-2}), we get
\begin{equation}
\frac{df_j}{dt} (q_i,p_i,t) = \left\{f_j,H \right\} +
\frac{\partial f_j}{\partial t} = 0, \label{invariant-condition}
\end{equation}
which means that the generators $f_j (q_i,p_i,t)$ are the
conserved quantities of the given physical system. Hence there is
a closed relation between the existence of a conserved quantity
and its associated symmetry canonical transformation. Since the
generators $f_j (q_i,p_i,t)$ are independent functions due to
independent parameters $\epsilon_j$, the conserved quantities
obtained must be independent to each other.

We may obtain the conserved quantities $f_j (q_i,p_i,t)$
associated with a given physical system by solving
Eq.~(\ref{invariant-condition}), which lead to the constructions
of its associated symmetry canonical transformations according to
Eq.~(\ref{infinitesimal-transformation}). Since
Eq.~(\ref{invariant-condition}) is a linear partial differential
equation of the first order, the following theorem of partial
differential equations provides the general solutions of the
equation~\cite{Dennemeyer68,Sneddon64} :
\begin{theorem}
Consider a linear partial differential equation of the first order
in the form of
\begin{equation}
\sum_{i=1}^{n} P_i (x_1, ... , x_n) \frac{\partial z}{\partial
x_i} = R (x_1, ... , x_n,z). \label{PDE-1}
\end{equation}
If $u_i(x_1, ... ,x_n,z) = c_i$, $i =  1 \, ... \, n$ are
independent solutions of the {\it subsidiary equation}
\begin{equation}
\frac{dx_1}{P_1} = \frac{dx_2}{P_2} = ... = \frac{dx_n}{P_n} =
\frac{dz}{R}, \label{sub-equation}
\end{equation}
then the relation $\Phi (u_1, ... ,u_n) = 0$ (where $\Phi$ is
arbitrary function) is a general solution of Eq.~(\ref{PDE-1}).
\label{theorem-PDE}
\end{theorem}
A rigorous proof of Theorem~\ref{theorem-PDE} is given in
Ref.~\cite{Sneddon64}. It should be noted that there are a {\it
maximum} of $n$ independent solutions satisfying
Eq.~(\ref{sub-equation}).

As the consequences of Theorem~\ref{theorem-PDE}, the general
solution of Eq.~(\ref{invariant-condition}) may be written as
\begin{equation}
f_j (q_i,p_i,t) = \tilde{\Phi} (u_1, ... ,u_{2n}),
\label{conserved-solution}
\end{equation}
where $n$ is the number of configuration dimension, $\tilde{\Phi}$
is an arbitrary function, and $u_k(q_i,p_i,t)$, $k = 1,2, ... ,2n$
are the independent solutions of
\begin{equation}
\frac{dq_1}{\left(\frac{\partial H}{\partial p_1} \right)} = ... =
\frac{dq_n}{\left(\frac{\partial H}{\partial p_n} \right)} = -
\frac{dp_1}{\left(\frac{\partial H}{\partial q_1} \right)} = ... =
-\frac{dp_n}{\left(\frac{\partial H}{\partial q_n} \right)} = dt.
\label{sub-equation-2}
\end{equation}
The form of the solution (\ref{conserved-solution}) indicates that
there could be infinitely many solutions satisfying
Eq.~(\ref{invariant-condition}).

Inserting (\ref{conserved-solution}) into
Eq.~(\ref{invariant-condition}), we have
\begin{equation}
\frac{du_k}{dt} (q_i,p_i,t) = \left\{u_k,H \right\} +
\frac{\partial u_k}{\partial t} = 0, \hspace{10mm} k = 1,2, ...
,2n. \label{invariant-condition-u}
\end{equation}
This means that {\it each independent solution} of the subsidiary
equation~(\ref{sub-equation-2}) corresponds to a conserved
quantity of the associated physical system. The maximum number of
independent solutions may be interpreted as the maximum number of
all independent conserved quantities one can possibly find in the
system, which depends on the number of configuration degrees of
freedom.

\section{Time-independent Conserved Quantities.}
\label{time-independent}

For the time-independent case, Eq.~(\ref{invariant-condition})
reduces to
\begin{equation}
\frac{df}{dt} (q_i,p_i) = \left\{f,H \right\} = 0.
\label{invariant-condition-2}
\end{equation}
The general solution of (\ref{invariant-condition-2}) may then be
written in the form of
\begin{equation}
f (q_i,p_i) = \bar{\Phi} (u_1, ... ,u_{2n-1}),
\label{conserved-solution-time-independent}
\end{equation}
where $u_i$, $i = 1 ... (2n-1)$ are the independent solution of
the subsidiary equation
\begin{equation}
\frac{dq_1}{\left(\frac{\partial H}{\partial p_1} \right)} = ... =
\frac{dq_n}{\left(\frac{\partial H}{\partial p_n} \right)} = -
\frac{dp_1}{\left(\frac{\partial H}{\partial q_1} \right)} = ... =
-\frac{dp_n}{\left(\frac{\partial H}{\partial q_n} \right)}.
\label{sub-equation-time-independent}
\end{equation}
It is clear from Eq.~(\ref{sub-equation-time-independent}) that it
is not possible to obtain a time-independent conserved quantity
from a time-dependent physical system, since the $\frac{\partial
H}{\partial p_i}$'s and $\frac{\partial H}{\partial q_i}$'s will
contain explicit time-dependence. The maximum number of possible
independent conserved quantities is $2n-1$. It should be noted
that Eq.~(\ref{invariant-condition-2}) admits a trivial solution
$f(q_i,p_i) = H(q_i,p_i)$, and hence $f(q_i,p_i) =
\Phi(H(q_i,p_i))$. This kind of solution will be automatically
satisfied in any time-independent case.

For the purpose of demonstration, we shall use
Eqs.~(\ref{conserved-solution-time-independent}) and
(\ref{sub-equation-time-independent}) to re-derive
time-independent conserved quantities of some well-known cases in
one- and many-particle systems.

\subsection{One-Particle Systems : Free Particles.}
\label{free-particle-system}

The simplest one-particle system is the {\it free particle
system}, whose Hamiltonian in 3-dimensional case is given by
\begin{equation}
H(q,p) = \frac{p_x^2 + p_y^2 + p_z^2}{2m}.
\label{3-free-hamiltonian}
\end{equation}
The correspondent subsidiary equation can be written
as
\begin{equation} \frac{dq_x}{\left(\frac{p_x}{m} \right)} =
\frac{dq_y}{\left(\frac{p_y}{m} \right)} =
\frac{dq_z}{\left(\frac{p_z}{m} \right)} = - \frac{dp_x}{0} = -
\frac{dp_y}{0} = \frac{dp_z}{0},
\label{subsidiary-3-free-particle}
\end{equation}
and its independent solutions are
\begin{eqnarray}
p_x & = & C_1, \nonumber \\
p_y & = & C_2, \nonumber \\
p_z & = & C_3, \nonumber \\
L_x & \equiv & q_yp_z - q_zp_y = C_4, \nonumber \\
L_y & \equiv & q_zp_x - q_xp_z = C_5.
\label{sub-solution-3-free-particle}
\end{eqnarray}
The general solution of Eq.~(\ref{invariant-condition-2}) of the
system is then
\begin{equation}
f(q,p) = \bar{\Phi} \left(p_x, p_y, p_z, L_x, L_y \right).
\label{general-solution-3-free-particle}
\end{equation}
From Eq.~(\ref{sub-solution-3-free-particle}), we have 5
independent solutions correspond to momenta in the $x$, $y$ and
$z$ directions, as well as angular momenta in the $x$ and $y$
directions. The angular momentum in the $z$ direction $L_z \equiv
q_xp_y - q_yp_x$ can be expressed as the function of 5 independent
conserved quantities as given in
Eq.~(\ref{general-solution-3-free-particle}).

\subsection{One-Particle Systems : Central Forces.}
\label{central-force-system}

The Hamiltonian of a central force system is given by
\begin{equation}
H(q,p) = \frac{p^2}{2m} + V(r),
\label{central-force-hamiltonian}
\end{equation}
where $V(r)$ is the central force potential. For the 3-dimensional
case, it is defined as
\begin{equation}
V(r) = V(q_x^2 + q_y^2 + q_z^2).
\label{central-force-potential}
\end{equation}

The associated subsidiary equation associated is given by
\begin{eqnarray}
&  & \frac{dq_x}{\left(\frac{p_x}{m} \right)} =
\frac{dq_y}{\left(\frac{p_y}{m} \right)} =
\frac{dq_z}{\left(\frac{p_z}{m} \right)} = -
\frac{dp_x}{\left(\frac{\partial V}{\partial q_x}\right)} = -
\frac{dp_y}{\left(\frac{\partial V}{\partial q_y}\right)} = -
\frac{dp_z}{\left(\frac{\partial V}{\partial q_z}\right)}
\nonumber \\
& \Longleftrightarrow & \frac{dq_x}{\left(\frac{p_x}{m} \right)} =
\frac{dq_y}{\left(\frac{p_y}{m} \right)} =
\frac{dq_z}{\left(\frac{p_z}{m} \right)} = - \frac{dp_x}{\left(2
q_x \frac{dV}{du}\right)} = - \frac{dp_y}{\left(2 q_y
\frac{dV}{du}\right)} \nonumber \\
&  & = - \frac{dp_z}{\left(2 q_z \frac{dV}{du}\right)},
\label{subsidiary-3-central-force}
\end{eqnarray}
where $u = q_x^2 + q_y^2 + q_z^2$. The independent solutions are
\begin{eqnarray}
H(q,p) & = & \frac{p_x^2 + p_y^2 + p_z^2}{2m} + V(q_x^2 + q_y^2 +
q_z^2) = C_1,
\nonumber \\
L_x & = & q_yp_z - q_zp_y = C_2, \nonumber \\
L_y & = & q_zp_x - q_xp_z = C_3, \nonumber \\
L_z & = & q_xp_y - q_yp_x = C_4.
\label{independent-3-central-force}
\end{eqnarray}
There are 4 independent solutions associated with an arbitrary
3-dimensional central force system, i.e. the trivial solution
associated with the total energy of the system, and 3 non-trivial
solutions associated with the angular momentum in $x$, $y$ and $z$
directions. However, there may be 5 independent solutions obtained
in some 3-dimensional systems, for instance, the 3-dimensional
harmonic oscillators. The general solution of
Eq.~(\ref{invariant-condition-2}) for the 3-dimensional central
force system is then
\begin{equation}
f(q,p) = \bar{\Phi} \left(H(q,p), L_x, L_y, L_z \right).
\label{general-solution-3-central-force}
\end{equation}

\subsection{Many-Particle Systems.}
\label{many-body-system}

We now consider the simplest case in many-particle systems.
Suppose there are $N$ objects with the same masses $m$. All the
objects lie on a straight line, and each object is connected with
its closest neighbourhoods by springs with constant $k$. For the
simplicity, let us consider only vibrations along the straight
line formed by the objects. The Hamiltonian of the system above is
given by
\begin{equation}
H(q_i, p_i) = \frac{p_1^2}{2m} + \sum_{i=2}^{N} \left[
\frac{p_i^2}{2m} + \frac{(q_i - q_{i-1})^2}{2} \right],
\label{hamiltonian-many-body}
\end{equation}
where $q_i$ and $p_i$ are conjugate coordinate and momentum of the
$i-th$ object. The associated subsidiary equation is
\begin{eqnarray}
&  & \frac{dq_1}{\left(\frac{p_1}{m} \right)} = ... =
\frac{dq_i}{\left(\frac{p_i}{m} \right)} = ... =
\frac{dq_N}{\left(\frac{p_N}{m} \right)} \nonumber \\
& = & \frac{dp_1}{k(q_2 - q_1)} = \frac{dp_2}{k(q_3 - 2q_2 + q_1)}
= ... \nonumber \\
& = & \frac{dp_i}{k(q_{i+1} - 2q_i + q_{i-1})} = ... =
\frac{dp_{N-1}}{k(q_N - 2q_{N-1} + q_{N-2})} \nonumber \\
& = & -\frac{dp_N}{k(q_N - q_{N-1})}. \label{subsidiary-many-body}
\end{eqnarray}
whose independent solutions are
\begin{eqnarray}
H(q_i, p_i) & = & \frac{p_1^2}{2m} + \sum_{i=2}^{N} \left[
\frac{p_i^2}{2m} + \frac{(q_i - q_{i-1})^2}{2} \right] = C_1,
\nonumber \\
\sum_{i=1}^{N} p_i & = & C_2, \nonumber \\
\sum_{i=1}^{N} q_i & = & C_3. \label{solution-many-body}
\end{eqnarray}
There are 3 independent solution associated with the system,
consisted of one trivial solution corresponds to the total energy
of the system, one corresponds to the conservation law of linear
momentum, and one solution which is equivalent to the conservation
of linear momentum since $p_i = m \dot{q}_i$. The general solution
of Eq.~(\ref{invariant-condition-2}) will be an arbitrary function
of $H(q,p)$, $\sum p_i$ and $\sum x_i$.

\section{Time-dependent Conserved Quantities.}
\label{time-dependent}

In the time-dependent case, Eq.~(\ref{invariant-condition}) must
be fully satisfied. There are two kinds of conserved quantities :
one associated with a time-independent Hamiltonian, and another
one associated with a time-dependent one. With the
time-independent Hamiltonian, Eq.~(\ref{invariant-condition})
still admits the trivial solution $H(q,p)$, which means the total
energy of the physical system is still conserved, as well as
time-independent solutions. With the time-dependent Hamiltonian,
this is no longer the case.

The subsidiary equation of Eq.~(\ref{invariant-condition}) is
given by Eq.~(\ref{sub-equation-2}), with the maximum number of
its independent solutions being $2n$. We shall use
Eqs.~(\ref{conserved-solution}) and (\ref{sub-equation-2}) to
derive time-dependent conserved quantities of time-independent
Hamiltonians, with one-particle systems as the examples, as well
as time-dependent conserved quantities associated with a
time-dependent Hamiltonian. We shall only discuss the
1-dimensional cases for the non-zero potential systems since
calculation difficulties arise in 2- and 3-dimensional as well as
in many-particle cases.

\subsection{Time-independent Physical Systems : Free Particles}
\label{time-independent-physical-system-free}

Using the Hamiltonian of 3-dimensional free particle system
(\ref{3-free-hamiltonian}), we write the subsidiary equation as
\begin{equation}
\frac{dq_x}{\left(\frac{p_x}{m} \right)} =
\frac{dq_y}{\left(\frac{p_y}{m} \right)} =
\frac{dq_z}{\left(\frac{p_z}{m} \right)} = - \frac{dp_x}{0} = -
\frac{dp_y}{0} = \frac{dp_z}{0} = dt.
\label{sub-3-free-particle-time-dependent}
\end{equation}
The independent solutions for the system are
\begin{eqnarray}
p_x & = & C_1, \nonumber \\
p_y & = & C_2, \nonumber \\
p_z & = & C_3, \nonumber \\
q_x - \frac{p_x}{m} t & = & C_4, \nonumber \\
q_y - \frac{p_y}{m} t & = & C_5, \nonumber \\
q_z - \frac{p_z}{m} t & = & C_6
\label{sub-solution-3-free-particle-time-dependent}
\end{eqnarray}
The last 3 solutions in
Eq.~(\ref{sub-solution-3-free-particle-time-dependent}) are the
expressions of homogeneous linear motions in 3-dimensional space.
The general solutions of Eq.~(\ref{invariant-condition}) for the
3-dimensional free particle systems are then
\begin{equation}
f(q,p,t) = \tilde{\Phi} \left(p_x, p_y, p_z, q_x - \frac{p_x}{m}
t,q_y - \frac{p_y}{m} t,q_y - \frac{p_y}{m} t \right),
\label{general-solution-3-free-particle-time-dependent}
\end{equation}
The angular momenta in 3-dimensional cases may be expressed by
(\ref{general-solution-3-free-particle-time-dependent}).

\subsection{Time-independent Physical Systems : Central Forces}
\label{time-independent-physical-system-central}

From the Hamiltonian of 1-dimensional free particle system
\begin{equation}
H(q,p) = \frac{p^2}{2m} + V(q^2), \label{Hamiltonian-free-1-D}
\end{equation}
one may write the associated subsidiary equation for
1-dimensional central force system as
\begin{equation}
\frac{dq}{\left(\frac{p}{m} \right)} = -
\frac{dp}{\left(\frac{dV}{dq}\right)} = dt.
\label{sub-equation-1-central-time-dependent}
\end{equation}

Unlike the derivations discussed in either
Section~\ref{time-independent} or
Section~\ref{time-independent-physical-system-free}, the treatment
of obtaining time-dependent solution of
Eq.~(\ref{sub-equation-1-central-time-dependent}) depends on the
{\it functional forms} of the central force potentials. For the
purpose of demonstration, we shall use the 1-dimensional harmonic
oscillator and the 1-dimensional inverse square potential system.

For the 1-dimensional harmonic oscillator with potential $V(q^2) =
\frac{1}{2} k q^2$, where $k =$ constant,
Eq.~(\ref{sub-equation-1-central-time-dependent}) becomes
\begin{equation}
\frac{dq}{\left(\frac{p}{m} \right)} = - \frac{dp}{k q} = dt,
\label{sub-equation-1-harmonic-time-dependent}
\end{equation}
which is equivalent to
\begin{equation}
\frac{d(p + i \sqrt{k m} \, q)}{p + i \sqrt{k m} \, q} = i
\sqrt{\frac{k}{m}} \, dt.
\label{equiv-sub-equation-1-harmonic-time-dependent}
\end{equation}
The independent solutions of
(\ref{sub-equation-1-harmonic-time-dependent}) are
\begin{eqnarray}
\frac{p^2}{2m} + \frac{1}{2} k q^2 = H(q,p) = C_1, \nonumber
\\
\ln(p + i \sqrt{k m} \, q) - i \sqrt{\frac{k}{m}} \, t = C_2,
\label{sub-solution-1-harmonic-time-dependent}
\end{eqnarray}
where the second independent solution is obtained by integrating
Eq.~(\ref{equiv-sub-equation-1-harmonic-time-dependent}). The
second solution of
Eq.~(\ref{sub-solution-1-harmonic-time-dependent}) gives us
information of how the momentum $p$ and coordinate $q$ behave in
the system. The general solution of (\ref{invariant-condition})
for a 1-dimensional harmonic oscillator is
\begin{equation}
f(q,p,t) = \bar{\Phi} \left(H(q,p),\ln(p + i \sqrt{k m} \, q) - i
\sqrt{\frac{k}{m}} \, t\right).
\label{general-solution-1-harmonic-time-dependent}
\end{equation}

With the 1-dimensional inverse square potential system, we use
$V(q^2) = -\frac{k}{q^2}$, where $k =$ constant.
Eq.~(\ref{sub-equation-1-central-time-dependent}) becomes
\begin{equation}
\frac{dq}{\left(\frac{p}{m} \right)} = -\frac{dp}{\frac{2k}{q^3}}
= dt. \label{sub-equation-1-inverse-time-dependent}
\end{equation}
The independent solutions of
Eq.~(\ref{sub-equation-1-inverse-time-dependent}) are
\begin{eqnarray}
\frac{p^2}{2m} - \frac{k}{q^2} = H(q,p) = C_1, \nonumber
\\
pq - 2 H(q,p) = C_2, \label{sub-solution-1-inverse-time-dependent}
\end{eqnarray}
where the second solution in
Eq.~(\ref{sub-solution-1-inverse-time-dependent}) is obtained from
the equivalent relation extracted from
Eq.~(\ref{sub-equation-1-inverse-time-dependent}) :
\begin{eqnarray}
p \, dq - \frac{p^2}{m} \, dt & = & 0, \nonumber \\
q \, dp + \frac{2k}{q^2} \, dt & = & 0.
\label{sub-equation-equiv-1-inverse-time-dependent}
\end{eqnarray}
The general solution of (\ref{invariant-condition}) for a
1-dimensional inverse square potential system is
\begin{equation}
f(q,p,t) = \bar{\Phi} \left(H(q,p), pq - 2 H(q,p), t\right).
\label{general-solution-1-inverse-time-dependent}
\end{equation}

\subsection{Time-dependent Physical Systems}
\label{time-dependent-physical-system}

As mentioned in the beginning of Section~\ref{time-dependent}, in
the time-dependent physical systems the Hamiltonians are no longer
conserved quantities, meaning that the subsidiary equations
Eq.~(\ref{sub-equation-2}) do not have the Hamiltonians as one of
their solutions.

Given the Hamiltonian of a 1-dimensional physical system as
\begin{equation}
H(q,p,t) = \frac{p^2}{2m} + V(q,t),
\label{time-dependent-hamiltonian}
\end{equation}
the subsidiary equations ~(\ref{sub-equation-2}) may be expressed
as
\begin{equation}
\frac{dq}{\left(\frac{p}{m} \right)} = -
\frac{dp}{\left(\frac{\partial V}{\partial q}\right)} = dt,
\label{sub-equation-1-time-dependent}
\end{equation}
where we have used partial derivative of $V(q,t)$ instead of total
derivative as in
Eq.~(\ref{sub-equation-1-central-time-dependent}).

As in the central force case in
Section~{\ref{time-independent-physical-system-central}, the
treatment of obtaining solutions for
Eq.~(\ref{sub-equation-1-time-dependent}) depends on the
functional forms of the associated potentials. For the purpose of
demonstration, let us consider a simple 1-dimensional system of a
point mass $m$ attached to a spring of constant $k$, where the
other end of the spring is fixed on a massless cart moving
uniformly along $q$ axis with speed $v_0$\footnote{The same
problem is also discussed in Ref~\cite[pp. 350]{Goldstein80} using
a different approach.}. The potential of the system is given by
\begin{equation}
V(q,t) = \frac{k}{2}(q - v_0 t)^2. \label{potential-cart}
\end{equation}
Inserting potential~(\ref{potential-cart}) into
Eq.~(\ref{sub-equation-1-time-dependent}), we have
\begin{equation}
\frac{dq}{\left(\frac{p}{m} \right)} = - \frac{dp}{k (q - v_0 t)}
= dt, \label{sub-equation-1-cart-time-dependent}
\end{equation}
which maybe equivalently written as
\begin{eqnarray}
\frac{p}{m} \, dp + k(q - v_0 t) \, dq = 0, \nonumber \\
dq - \frac{p}{m}\, dt = 0, \nonumber \\
dp + k(q - v_0 t) \, dt = 0.
\label{sub-equation-1-cart-time-dependent-equiv}
\end{eqnarray}
Applying the total derivative of $V(q,t)$ to the first equation of
(\ref{sub-equation-1-cart-time-dependent-equiv}) and using the
third equation of
(\ref{sub-equation-1-cart-time-dependent-equiv}), we get
\begin{equation}
\frac{p}{m} \, dp + dV - v_0 \, dp = 0,
\end{equation}
which can be easily integrated to obtained the associated
independent solution :
\begin{eqnarray}
&   & \frac{p^2}{2m} + \frac{1}{2} k (q - v_0 t)^2 - v_0 p = C,
\nonumber \\
& \Longleftrightarrow & \frac{(p - mv_0)^2}{2m} + \frac{1}{2} k (q
- v_0 t)^2 - \frac{1}{2} m v_{0}^2 = C, \nonumber \\
& \Longleftrightarrow & \frac{(p - mv_0)^2}{2m} + \frac{1}{2} k (q
- v_0 t)^2 = C_1.
\label{sub-equation-1-cart-time-dependent-solution}
\end{eqnarray}
Although the independent solution has the dimension of energy, it
is {\it not} the total energy of the associated system, but the
total energy of the point mass relative to the cart instead. The
general solution of (\ref{invariant-condition}) for the associated
system is
\begin{equation}
f(q,p,t) = \bar{\Phi} \left(\frac{(p - mv_0)^2}{2m} + \frac{1}{2}
k (q - v_0 t)^2 \right). \label{general-solution-1-cart}
\end{equation}

\section{Summary and Conclusion.}
\label{summary}

We have discussed the way of obtaining conserved quantities of a
given physical system by solving Eq.~(\ref{invariant-condition})
using the subsidiary equations~(\ref{sub-equation-2}). The maximum
number of possible conserved quantities a physical system may have
is given by Theorem~\ref{theorem-PDE}. Some particular
time-independent cases for one-particle and many-particle systems,
as well as time-dependent cases for one-particle system have been
used to demonstrate the technique and treatment of handling
Eq.~(\ref{sub-equation-2}). For the same physical systems, it is
found that one may not have the same results for the
time-independent and time-dependent cases. More complicated
treatment may be used to deal with time-dependent cases and
many-particle systems.

Although the technique presented here is in the scope of classical
mechanics, it may be applicable to classical fields, provided that
the Poisson brackets in Eq.~(\ref{invariant-condition})
re-formulated in an appropriate way in terms of field
variables~\cite{Sudarshan74,Goldstein80}. Also, since the
connection between classical and quantum mechanics can be
established by replacing the Poisson brackets with Lie brackets or
commutators, the conserved quantities in quantum mechanics may be
obtained directly by replacing $(q,p)$ in the associated classical
quantities with $(q,\frac{\hbar}{i} \, \frac{\partial}{\partial
q})$.

As the symmetries of the physical system yields the existence of
non-trivial conserved quantities in classical mechanics, the same
argument is true for quantum system. Moreover, the existence of
symmetries in quantum mechanics generates degeneracy which depends
on the number of degree of freedom. For example, since there is no
non-trivial conserved quantity in a time-independent one
dimensional harmonic oscillator, no degeneracy occurs in the
quantum counterpart. The degeneracy of harmonic oscillator system
shows up in 2- and 3-dimensional cases.

\section*{Acknowledgement}
The authors would like to thank Prof. B. Suprapto Brotosiswojo and
Dr. A. Rusli of the Department of Physics, Institut Teknologi
Bandung - Indonesia for their helpful discussions and valuable
comments concerning the subject.


\section*{References.}
\thebibliography{20}

\bibitem{Sudarshan74} Sudarshan, E.C.G. \& N. Mukunda., {\it Classical Dynamics : A Modern Perspective},
John Wiley and Sons., New York, 1974.
%
\bibitem{Itzykson80} Itzykson, C. \& Zuber, J., {\it Quantum Field
Theory}, McGraw-Hill Inc., New York, 1980.
%
\bibitem{Weinberg95} Weinberg, S., {\it Quantum Theory of Fields},
Cambridge University Press, New York, 1995.
%
\bibitem{Abers73} E. Abers and B. Lee, Phys. Rep. {\bf 9}C, 2
(1973).
%
\bibitem{Cheng84} Cheng, Ta-Pei \& Ling-Fong Li, {\it Gauge theory
of elementary particle physics}, Clarendon Press, Oxford, 1984.
%
\bibitem{Moriyasu85} Moriyasu, K., {\it An Elementary Premier for
Gauge Theory}, World Scientific, Singapore, 1985.
%
\bibitem{Goldstein80} Goldstein, H., {\it Classical Mechanics}, Addison-Wesley, Massachusetts, 1980.
%
\bibitem{Pauri66} M. Pauri and G.M. Prosperi., J. Math. Phys. {\bf 7}, 366
(1966).
%
\bibitem{Sutanto91} Sutanto, S. H., {\it Algebraic Structures in
Canonical Formalisms}, BSc. thesis, Department of Physics, Bandung
Institute of Technology, Indonesia, 1991.
%
\bibitem{Gilmore74} Gilmore, R., {\it Lie Groups, Lie Algebras and Some of Their Applications}
, John Wiley and Sons, New York, 1974.

\bibitem{Dennemeyer68} Dennemeyer, R., { \it Introduction to
Partial Differential Equations and Boundary Value Problems},
McGraw-Hill Book Co. New York, 1968.

\bibitem{Sneddon64} Sneddon, Ian, {\it Elements of Partial
Differential Equations}, McGraw-Hill Book Co., New Delhi, 1964.
\end{document}